\documentclass[english,aps,prl,twocolumn,showpacs,groupedaddress,floatfix]{revtex4}

\usepackage{amsmath} 
\usepackage{bbm}
\usepackage{epsfig}
\usepackage{times}
\usepackage{babel}
\makeatletter

\newcommand{\id}{\mathbbm{1}}
\newcommand{\cc}{{\mathbbm{C}}}
\newcommand{\rr}{{\mathbbm{R}}}

\begin{document}

\title{Renormalization algorithm with graph enhancement}

\author{R.\ H\"ubener, C.\ Kruszynska,
L.\ Hartmann, and W.\ D\"ur}

\affiliation{Institut f\"ur Theoretische Physik, Universit\"at Innsbruck,
Technikerstra\ss e 25, A-6020 Innsbruck, Austria\\
 Institut f\"ur Quantenoptik und Quanteninformation der \"Osterreichischen
Akademie der Wissenschaften, Innsbruck, Austria}

\author{F.\ Verstraete}
\affiliation{Fakult\"at f\"ur Physik, Universit\"at Wien, Boltzmanngasse 5,
A-1090 Wien, Austria}

\author{J.\ Eisert and M.B.\ Plenio}
\affiliation{Institute for Mathematical Science, Imperial College London, London SW7 2PG, UK}
\affiliation{Blackett Laboratory, Imperial College London,
London SW7 2BW, UK}

\date{\today}

\begin{abstract}
We introduce a  class of variational states to describe quantum 
many-body systems. This class generalizes matrix product states 
which underly the density-matrix renormalization group approach 
by combining them with weighted graph states.  States within
this class may (i) possess arbitrarily long-ranged two-point 
correlations, (ii) exhibit an arbitrary degree of block entanglement 
entropy up to a volume 
law, (iii)  may be taken translationally invariant, while at 
the same time (iv) local properties and two-point correlations can 
be computed efficiently. This new variational class of states can 
be thought of as being prepared from matrix product states, followed 
by commuting unitaries on arbitrary constituents, hence truly 
generalizing both  matrix product  and weighted graph 
states. We use this class of states to formulate a renormalization 
algorithm with graph enhancement (RAGE) and present numerical 
examples demonstrating that improvements over density-matrix 
renormalization group simulations can be achieved in the simulation 
of ground states and quantum algorithms. Further generalizations, 
e.g., to higher spatial  dimensions, are outlined.
\end{abstract}


\maketitle

Strongly correlated quantum systems give rise to a number 
of intriguing phenomena in condensed matter systems such 
as the existence of rare-earth magnetic insulators or 
high-temperature superconductors. The classical description 
of such quantum many-body systems is difficult, as entanglement and 
interactions cannot be neglected. A full solution of the 
underlying microscopic model is unfeasible due to the 
exponential growth of the dimension of the Hilbert space 
with the number of constituent particles. In turn, numerical 
variational approaches, like the {\it density-matrix 
renormalization group} (DMRG) technique \cite{Wh91,Sc04}, 
make use of an important observation. Typically,
ground or thermal states do not occupy the exponentially 
large Hilbert space, but a much smaller subspace. DMRG
can indeed be seen as a variation over the polynomially
sized set of matrix product states (MPS) 
\cite{FNW92,Rommer95,Vi03,Ve04}, approximating the true 
ground state iteratively. This approach is expected to 
work particularly well in one-dimensional gapped systems, 
in which correlation functions decay exponentially and 
the entanglement entropy saturates at 
larger block sizes, satisfying an ``area law'' \cite{Area}.

Any such variational method, however, has its limitations. 
For example, in a critical one-dimensional system, the MPS 
description is no longer economical, with other variational 
sets potentially being more appropriate. When it comes to 
time evolution, area laws may be replaced by volume laws 
\cite{Calabrese}, and a DMRG picture can become very expensive. 
{\it Projected entangled pair states} (PEPS) form 
higher-dimensional analogues of {\it matrix product states}
MPS \cite{Ve04b}. This approach is very promising but still 
in development. For critical systems, {\it multi-scale entanglement
renormalization} or {\it contractor renormalization} \cite{MERA,Flow}
are promising candidates also in two dimensions, 
but are not easily reconcilable with translational invariance.
{\it Weighted graph states} (WGS) \cite{Du05,Ca05,Weighted} 
are a family of states that can embody long-range correlations
in any spatial dimensions, but do not seem to grasp short-range 
properties as well as MPS do \cite{Weighted1}. 

With these observations in mind, one of the key questions
seems to be the following. How far can one go with the 
efficient classical description of quantum many-body systems? 
Can MPS for example be generalized to a larger class of 
states encompassing some of the above approaches while 
retaining all of their convenient features? Can one have 
additional long-range correlations while still being able 
to efficiently compute local properties and correlation functions? 
In particular, given the complementary strengths of the 
MPS and the WGS approach it is natural to attempt a 
unification of the two approaches. This work shows that 
indeed the two pictures can be combined to form a new 
enlarged variational set, while retaining all of the 
desirable structural elements of its ancestors. We first 
define the set, discuss variations, sketch generalizations 
and finally demonstrate applicability and performance 
as well as limitations in ground state approximations 
and simulations of quantum algorithms. 

{\it Renormalization algorithm with graph enhancement. --}
We start from MPS of a quantum chain of length
$N$, consisting of $d$-level systems, as 
used in DMRG \cite{FNW92,Rommer95,Vi03,Ve04}  
\begin{equation}
        |\psi(A)\rangle := \sum_{s_1,\ldots,s_N=0}^{d-1}
        \text{tr} [A_{s_1}^{(1)}\ldots A_{s_N}^{(N)} ]
        |s_1,\ldots,s_N\rangle
\end{equation}
where the $A_{s_n}^{(n)}$ are complex $D\times D$ matrices. 
For open boundary 
conditions, the left- and rightmost matrices can be taken 
to be vectors. For simplicity of notation, but in a way 
that can be trivially generalized, we now fix $d=2$.
% for a spin chain. 
MPS have correlation
functions $\langle Z^{(j)} Z^{(j+k)} \rangle
- \langle Z^{(j)} \rangle\langle Z^{(j+k)}\rangle$ exponentially decaying in $k$ 
and satisfy an area law \cite{Area} by construction 
\footnote{In our notation the the brackets in the upper position
denote the support of an operator.}.
An area law in 1D implies that any Renyi entropy $S_\alpha$ 
of the reduced state of a block of $L$ contiguous spins will 
eventually saturate ($S_\alpha(\rho_L)= O(1)$);
many ground states possess this property 
%If the true ground state satisfies such a law, 
and hence a good and economical MPS 
approximation of them is possible \cite{Schuch}. 

Now we go beyond this picture and apply to the MPS 
any set of commuting unitaries between any two constituents,
irrespective of the distance. More specifically, 
we consider the adjacency matrix $\Phi$ of a weighted 
simple graph with $\Phi_{k,l}\in [0,2\pi)$ 
and apply without loss of generality the corresponding {\it phase gates} 
$U(\Phi_{k,l}):= |0,0\rangle\langle 0,0|+ |0,1\rangle\langle 0,1|+
|1,0\rangle\langle 1,0|+ |1,1\rangle\langle 1,1|e^{i \Phi_{k,l}}$ 
between the particles $k,l$ in the chain.
%connected by an edge of the graph. 
Finally, we apply local rotations $V_j\in U(2)$, to arrive at the variational 
class of states defined by
\begin{multline}
    |\psi(A,\Phi,V)\rangle :=\prod_{j=1}^N V_j^{(j)} 
    \prod_{k,l} U^{(k,l)}(\Phi_{k,l})\\
    \times
     \sum_{s_1, \ldots , s_N} 
     \text{tr}
    [A_{s_1}^{(1)}\ldots A_{s_N}^{(N)}]
    |s_1, \ldots , s_N\rangle,\label{RAGE}
\end{multline}
which then forms the basis of the {\it renormalization group 
algorithm with graph enhancement} (RAGE).
%
%{\it Relationship with matrix product and weighted graph states. --}
The above set clearly embodies a large variational class.
By definition, for $\Phi=0$ and $V_j=\id$, it includes the 
MPS. It also includes superpositions of WGS as first considered 
in Ref.\ \cite{Weighted}, 
\begin{eqnarray}
    |\varphi\rangle\!\!
    &=& \!\! \sum_m \alpha_m \prod_{j=1}^N V_j^{(j)}\!\!\!\!\!\!\!
    \sum_{s_1,\ldots,s_N=0}^1
     e^{-i  \mathbf{s}^T\Phi \mathbf{s}+  \mathbf{d}_m^T \mathbf{s}}
    |s_1,\ldots ,s_N\rangle  
    \label{defwgraph}\\
    &=& \prod_{j=1}^N V_j^{(j)} \prod_{k,l} U^{(k,l)}(\Phi_{k,l})
    \sum_m \alpha_m  |\eta_{m,1}\rangle\otimes
    \ldots\otimes|\eta_{m,N}\rangle\nonumber
\end{eqnarray}
where $\mathbf{d}_m=(d_{m,1},\dots, d_{m,N})$, $\mathbf{s}=(s_1,\dots, s_N)$,
$|\eta_{m,n}\rangle := |0\rangle + e^{d_{m,n}} |1\rangle$
and $U(\Phi_{m,n})$ are defined as above, and which
can be shown to be of the form of Eq.\ (\ref{RAGE}).
For simplicity, and w.l.o.g, we will often set $V_j=\id$ 
subsequently.

{\it Main properties of RAGE states. --} 
To start with, RAGE states have a polynomially sized description, 
where the MPS and the WGS part are fully determined by $O(ND^2)$ 
and $O(N^2)$ real parameters respectively. Furthermore

{\it (i) Volume law for the entanglement entropy:}
By having a collection of maximally entangled qubit pairs across a 
boundary, the von-Neumann entropy of a block of length $L$ can 
be taken to scale as $S(\rho_L) = O(L)$. Encompassing graph 
states, our class can hence maximize the entanglement 
entropy.

{\it (ii) Translational invariance:} Whenever the MPS part 
is translationally invariant, $\Phi$ is a cyclic matrix, and 
$V_j$ is the same for all $j$, the whole state $|\varphi\rangle$ is manifestly 
translationally invariant. There exist other translational
invariant states that do not have this simple form. The 
key feature, though, is that unlike for multiscale entanglement 
renormalization \cite{MERA}, there exists this natural subset 
of states for which translational invariance is guaranteed 
to be exactly fulfilled, while at the same time a volume 
law for block-wise entanglement is possible \cite{Du05,Ca05}.

{\it (iii) Completeness:} As MPS already form a complete 
set in Hilbert space (if one allows $D$ to scale as 
$O(2^N)$, one can represent any pure state in 
$({\cc}^2)^{\otimes N}$) and this remains true for the 
RAGE set.

{\it Efficient computation of local properties and correlation functions. --}
The previous properties are all very natural and desirable,
and especially (i) cannot be achieved  efficiently  
with MPS alone.  However, as will be shown, this does not
prevent us from computing local properties
and correlation functions efficiently -- which is the key feature of this set.

To compute expectation values of observables
with small support 
we use the relevant reduced density
matrix $\rho_{\cal S}$, which may be computed efficiently with an 
effort of $O(ND^5)$ in the total size $N$ of the system 
${\cal S}\subset \{1,\dots, N\}$. Controlled phase gates acting exclusively
on qubits that are traced out make no contribution, while 
those that act on the spins in ${\cal S}=\{m_1,\dots, m_{|{\cal S}|}\}$ 
amount to a redefinition of the observables that 
are local to ${\cal S}$. Therefore we redefine $\Phi$ such that the 
above simplifications
are enforced, $\omega_{k,l}=\Phi_{k,l} \text{ if } k\in{\cal S},l\in \bar{\cal S}$ 
or $k \in \bar {\cal S}, l \in {\cal S}$, where $\bar {\cal S}=\{1,\ldots,N\}\setminus{\cal S}$, and $\omega_{k,l}=0$ otherwise.
We also define $E^{(j)}_{k,l}:= A_{k}^{(j)}\otimes (A_{l}^{(j)})^\ast$,
where $\ast$ denotes complex conjugation.
The reduced density matrix $\rho_{\cal S}$ (up to phase gates in ${\cal S}$)
is then found to be
\begin{eqnarray*}
    \rho_{\cal S}&=&
    \sum_{{s_1,\ldots,s_N=0}\atop{r_1,\ldots,r_N=0}}^1
    \text{tr} [ E_{s_1,r_1}^{(1)}
    \ldots E_{s_N,r_N}^{(N)}
     ] \text{tr}_{\bar {\cal{S}}}
     [  (\prod_{k,l}
     U^{(k,l)}(\omega_{k,l}))\nonumber\\
     &&\times
     |s_1,\ldots ,s_N\rangle\langle r_1,\ldots , r_N|
    (\prod_{k,l} U^{(k,l)\dagger}(\omega_{k,l})) 
    ]\nonumber\\
    &=&
     \sum_{{s_1,\ldots,s_N=0}\atop{r_1,\ldots,r_N=0}}^1
    \text{tr} [ E_{s_1,r_1}^{(1)}
    \ldots E_{s_N,r_N}^{(N)}
     ] |s_{m_1},\dots, s_{m_{|\cal{S}|}}\rangle
     \nonumber\\
     &&\times \langle r_{m_1},\dots, r_{m_{|\cal{S}|}}| 
     \prod_{k\in {\cal{S}}, l \in{\bar {\cal{S}}}}
     e^{i \omega_{k,l}(\delta_{s_k,1} - \delta_{r_k,1})\delta_{s_l,1}}.
     \nonumber
\end{eqnarray*}
The key of the above argument is that the effect of 
the phases is a mere modification of the transfer 
operators of the MPS by a phase factor, the phase 
depending on the matrix element in question. Thus, 
the evaluation of expectation values is performed 
using (products of) transfer operators associated 
with the single sites. The reduced state can then 
be written as 
\begin{eqnarray*}
    \rho_{\cal S}&=&
   \sum_{
   {s_{m_1},\dots, s_{m_{|\cal S|}}=0}
   \atop
   {r_{m_1},\ldots,r_{m_{|\cal S|}}=0}}^1
    \text{tr} \bigl[ \prod_{k=1}^N T_{s_k,r_k}^{(k)}(\{ s_{m_p},r_{m_p}:m_p\in{\cal S} \} )
    \bigr]
    \\
    &\times&| s_{m_1},\dots, s_{m_{|\cal S|}}\rangle\langle r_{m_1},\dots, r_{m_{|\cal S|}}|,
\end{eqnarray*}
where now
% \begin{equation*}
        $T_{s_k,r_k}^{(k)}(\{ s_{m_p},r_{m_p}\}):=
        E^{(k)}_{s_k,r_k}$
%\end{equation*}
if $k\in {\cal S}$, 
which are the unmodified transfer operators of MPS, and 
 \begin{eqnarray*}
         T_{s_k,r_k}^{(k)}(\{ s_{m_p},r_{m_p}\}) &:=& \sum_{l=0}^1  
        B_{l}^{(k)}(\{s_{m_p}\},s_k)\nonumber\\
        &\otimes & (B_{l}^{(k)}(\{r_{m_p}\},r_k))^\ast     
\end{eqnarray*}
if $k\in {\bar {\cal S}}$, which are the transfer operators modified
by phases,
%\begin{eqnarray*}    
        $B_{l}^{(k)}(\{ s_{m_p}\},s_k):=
        A_{l}^{(k)}\prod_{m_p\in {\cal S}} 
     e^{i\omega_{m_p,k}\delta_{s_{m_p}, 1}\delta_{s_k,1}}$.
%\end{eqnarray*}
%Note that the transfer operators do not depend on all variables of ${\cal S}$ for $k\in {\cal S}$, and in turn do not depend on $r_k,s_k$ for $k\in {\bar {\cal S}}$. 
Grouped in this way, the reduced density 
operator can indeed be evaluated efficiently. In fact, for each 
$\{ r_{m_p},s_{m_p}\}$ the effort to compute the entry of 
the reduced state is merely $O(N D^5)$, 
as one has to multiply $N$ transfer matrices of dimension
$D^2\times D^2$, just as in the case of MPS. This procedure is
inefficient in $|{\cal S}|$, with an exponential scaling 
effort. However, any Hamiltonian with two-body (possibly 
long-ranged) interactions can be treated efficiently term by term.
%However, $|{\cal S}|$ depends on the support of the operator
%to be evaluated only, and for typical Hamiltonians $H=\sum_{<i,j>} H_{i,j}$
%this means $|{\cal S}|=2$, since we treat each summand separately.

{\it Efficient updates. --} Besides procedures for the efficient 
computation of reduced density matrices, and therefore expectation 
values, we need a variational principle to improve the trial 
states. We will focus on local variational 
approaches to approximate ground states by minimizing
the energy
\begin{equation}\label{Energy}
        E:=\frac{\langle \psi(A,\Phi,V) |H |  
        \psi(A,\Phi,V)\rangle}{\langle \psi(A,\Phi,V) |  
        \psi(A,\Phi,V)\rangle},
\end{equation}
on the approximation of time evolution and
on the simulation of quantum circuits. The search for ground 
states is well known to be related to imaginary-time 
evolution. 

{\it Static updates. --} The MPS part can
be updated as in variants of DMRG \cite{Ve04}. The
expression
$\langle \psi(A,\Phi,V) |H |  
        \psi(A,\Phi,V)\rangle$ is (as in MPS) 
        a quadratic form in each of the
        entries of the matrices $A^{(k)}_0$, $A^{(k)}_1$
        for each site $k=1,\dots, N$.
        An optimal local update can therefore
        be found by means of solving generalized
        eigenvalue problems with an effort
        of $O(D^3)$. Local rotations can be
        incorporated by parametrizing single 
qubit rotations on spin 
$k$ with real parameters $x_k\in \rr^4$ as
$V_k = \sum_{j=1}^4 x_{k,j} M_j$
with $M=(\id, \sigma_z, \sigma_y,\sigma_x)$ being the
vector of the Pauli matrices including the identity. Again,
the local variation of $x_k$ in (\ref{Energy}) is a generalized 
eigenvalue problem in $x_k$ for each site $k=1,\dots, N$.
To optimize the phases of the WGS, 
one may first define the new Hamiltonian
    $H_V := (\prod_{j=1}^N V_j^{(j)\dagger}) H (\prod_{k=1}^N V_k^{(k)})$.
The optimal phase gates between any pair of spins 
$j,k\in\{1,\dots,N\}$ can be computed efficiently
%from minimizing $\langle\psi (A,\phi,V)|P_{j,k}^{\dagger}H_V P_{j,k}|\psi(A,\phi,V)\rangle/
%\langle\psi (A,\phi,V) |\psi(A,\phi,V)\rangle$,
%where $P_{j,k}:=a\id+ib\sigma_z\otimes\sigma_z$ with 
%nontrivial support on $j$ and $k$, with $a,b\in \rr$
%is simple 
as the procedure amounts to a quadratic function of a 
single variable $z=e^{i\Phi_{j,k}}$. To summarize, an 
update of $|\psi(A,\Phi,V)\rangle$ to minimize (\ref{Energy}) 
corresponds to a sweeping over such local variations, 
each of which is efficiently possible, with an effort 
of $O(MND^3)$ for $M$ sweeps. An element that is not 
present for MPS alone is that one can make a choice
whether one adapts an MPS part or the adjacency
matrix for an identical change in the physical state.
In practice, we have supplemented this 
procedure with a gradient-based global 
optimization, making use of the fact that 
the gradient can be explicitly computed. 

\begin{figure}[ht]
%\hspace*{-1.4cm}\vspace*{-1.0cm}
\includegraphics[width=.49\textwidth]{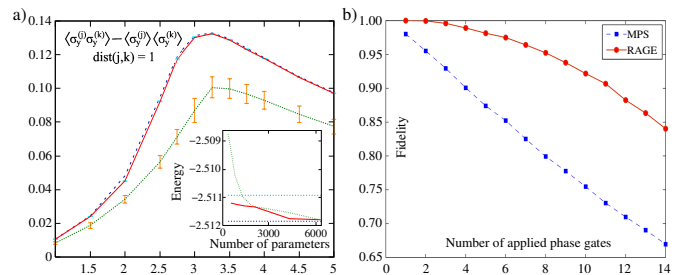}

\caption{(a) 2D Ising model on a  $4\times4$ periodic lattice . 
We compare the achievable accuracy with RAGE (red, solid) and MPS 
(green, dotted) with $D=4$ with exact results (blue, dashed). 
%The total number of independent parameters is 384 for MPS and 552 for the RAGE state.
Two-point correlations as function of $B$ are shown.
The inset depicts the energy for different {\em total} numbers of parameters and 
$B=2$ in comparison with the exact ground state (blue, dashed) as 
well as the first excited state (light blue, dashed). 
(b) Comparison of MPS (blue, dashed) and RAGE (red, solid) with $D = 2$ for
the simulation of a random quantum circuit \cite{Biham} on $N = 14$
qubits. Application of a random local phase gate followed by a 
random controlled-phase gate with random uniform phase
in $[0, 2\pi)$ constitutes one block. For given $k$ we apply this
block $k$ times to a randomly chosen initial 
MPS state. $500$ such runs are determined, and 
in each the fidelity with the exact
state is computed. The average over $500$ realizations is then
plotted.}\label{SimulationFigure}\label{Ising}
\end{figure}

We have applied the RAGE-method to proof-of-principle 
1D and 2D models, where the adjacency matrix is allowed 
to connect any constituents in the lattice. Fig.\ \ref{SimulationFigure}(a) 
shows results for the 2D Ising model with transversal magnetic 
field, $H=\sum_{\langle a,b\rangle }
\sigma_z^{(a)}\sigma_z^{(b)} + B\sum_a\sigma_x^{(a)}$,
comparing the achievable accuracy of MPS (using 
a one-dimensional path in the 2D lattice) and the 
RAGE-method for a fixed total number of free parameters.
The RAGE-method gives a significantly better accuracy 
regarding ground state energy and two-point correlations, 
already for a very small number of parameters. For other 
models, we see a similar improvement of RAGE over 
MPS, although in some cases (e.g., for a 2D Heisenberg 
model) the overall accuracy is still not very satisfactory,
possibly related to local minima encountered in the 
procedure. This new class of states 
does allow for new features like
long-range correlations and a violation of an area law, but
in turn, breaks the local $SU(2)$ gauge invariance. It is 
also clear from the simulations that the limitation of the 
underlying 1D structure of the MPS cannot always be fully 
overcome by the graph enhancement. The full potential in numerical
performance in identifying ground states is yet to be explored.
There exist, however, a number of interesting parent Hamiltonians 
where the RAGE method should be particularly well suited, e.g., perturbations 
of models which have a WGS as an exact or approximate ground 
state. We mention Kitaev's model (and perturbations 
thereof) on a hexagonal lattice which has the toric 
code state -- a WGS -- as ground state.

{\it Time evolution and simulation of quantum circuits. --} 
We have also considered time evolution, more specifically 
the evolution of a quantum state in a quantum circuit. Here, 
sequences of elementary gates are applied, e.g., two-qubit 
phase gates and arbitrary single-qubit rotations. This method 
can be easily adapted to Hamiltonian (real or imaginary) time 
evolution. We now show how to efficiently obtain an optimal 
approximation of the resulting state after the application 
of an elementary gate. It turns out to be useful to restrict 
the variational family by setting $V_j^{(j)}=\id$, although 
an extension to arbitrary $V$ is possible. 
For phase gates, this update is particularly simple, 
as only a change in the adjacency matrix $\Phi$ is required. 
It is part of the strength of the scheme that phase gates between 
arbitrary constituents are already
included in the variational set. The update of a 
local unitary will require some more attention: 

Consider an initial state vector
$|\psi(A,\Phi,\id)\rangle$, to which a single qubit 
unitary operation $U$ is applied -- acting, e.g., on the first qubit. 
The goal is now to find the best approximation $|\psi(A',\Phi',\id)\rangle$ 
which maximizes  
\begin{eqnarray}
    O:= \frac{|\langle \psi(A',\Phi',\id)|U_1 |\psi(A,\Phi,\id)\rangle|^2}
    {\langle \psi(A',\Phi',\id)|\psi(A',\Phi',\id)\rangle
    }.
    \label{inner}
\end{eqnarray}
It appears natural to vary only phases that directly affect 
qubit 1, i.e., $\Phi'_{j,k}= \Phi_{j,k}$ if $j\not=1$. In 
this case, one can rewrite (\ref{inner}) in such a way that 
the optimal MPS part $A'$ can be obtained analytically by 
solving a set of linear equations, while the optimization 
of a single phase $\Phi'_{1,k}$ leads to a simple quadratic 
form. In practice, an alternating sweeping of both kinds of 
local variational methods is required. We have tested this 
method for a random quantum circuit (see Fig.\ 
\ref{SimulationFigure}(b)) and compared the achievable 
accuracy with MPS. Again, we obtain an improvement due to 
the WGS. 
{\it Extensions. --}
A similar construction as illustrated for MPS also
works for unifying WGS with other underlying tensor network
descriptions. Similarly, one can use arbitrary clifford circuits
instead of the WGS and can still efficiently contract. 
More precisely, whenever an exact or approximate evaluation
of expectation values of arbitrary product observables (i.e., tensor
products of local operators)
for a state described by a tensor network is possible,
then \emph{local} observables (i.e., observables with a small support)
can be efficiently computed for the unified family of such tensor
network states and WGS (or clifford circuits), 
following an approach similar as in Eq.\
(\ref{defwgraph}). While this certainly restricts
the set of computable quantities (e.g., string-order parameters can
no longer be evaluated), it still suffices to compute expectation
values of all \emph{local} Hamiltonians and hence one obtains a variational
method for a ground-state approximation or simulation of quantum circuits.

{\it Conclusions. --}
To summarize, we have introduced a new variational class
of states to describe quantum many-body systems. These states
have a number of desirable properties. Correlation functions can be 
computed efficiently, systematic improvements of the approximation
within the class are possible and the states carry long-range 
correlations and violate entanglement area laws, as being 
encountered in critical systems or in quenched quantum systems 
undergoing time evolution. We have applied the RAGE ansatz
to condensed matter and quantum computation problems, where we 
find an improvement over MPS. From a fundamental perspective 
the key question is where 
exactly the boundaries for the efficient classical description of 
quantum systems might lie. In fact, intriguingly, the entanglement content of the 
state cannot be taken as an indicator for the ``complexity of 
a state'' \cite{Kr07}. Delineating this boundary will reveal 
more about the structure of quantum mechanics from a complexity 
point of view and holds the potential for new improved 
algorithms and methods for the description of quantum systems.

{\it Acknowledgements. --} We thank S.\ Anders,
T.J.\ Osborne and  C.M.\ Dawson for illuminating 
discussions and the FWF, the EU (QAP, OLAQUI, SCALA), 
the EPSRC QIP-IRC, the Royal Society, Microsoft 
Research, and a EURYI for support.


\begin{thebibliography}{99}
%
%\bibitem{qmc}
%D.~M.\ Ceperley, in \textit{The Monte Carlo Method in Physical Sciences},
%ed. J.~E.\ Gubernatis, AIP Conf.\ Proc.\ \textbf{690}, 85 (2003),
% arXiv physics/0306182
%
\bibitem{Wh91}
        S.R.\ White, Phys.\ Rev.\ Lett.\ {\bf 69}, 2863 (1992);
        Phys.\ Rev.\ B {\bf 48}, 10345 (1993).
%
\bibitem{Sc04}
        U.\ Schollw\"ock, Rev.\ Mod.\ Phys.\ {\bf 77}, 259 (2005).
%
\bibitem{FNW92}
        M.\ Fannes, D.\ Nachtergaele, and R.F.\ Werner, Commun.\ 
        Math.\ Phys.\ {\bf 144}, 443 (1992).
%
\bibitem{Rommer95}
        S.\ Rommer and S.\ Ostlund, 
        Phys. Rev. B {\bf 55}, 2164
        (1997); Phys. Rev. Lett. {\bf 75}, 3537 (1995).
        
\bibitem{Vi03}
        G.\ Vidal, Phys.\ Rev.\ Lett. {\bf 91}, 147902 (2003);
        G.\ Vidal, Phys.\ Rev.\ Lett. {\bf 93}, 040502 (2004).

\bibitem{Ve04}
        F.\ Verstraete, D.\ Porras and J.I.\ Cirac, 
        Phys. Rev. Lett. {\bf 93}, 227205 (2004).
        
\bibitem{Area}
        K.\ Audenaert, J.\ Eisert, M.B.\ Plenio, and R.F.\ Werner,
        Phys.\ Rev.\ A {\bf 66}, 042327 (2002);
%        G. Vidal, J. I. Latorre, E. Rico and A. Kitaev,  
%        Phys.\ Rev.\ Lett.\ {\bf 90}, 227902 (2003);
        I.\ Peschel, %On the entanglement entropy for an XY spin chain. 
        J.\ Stat. Mech.: Th. Exp. P12005 (2004);
        A.R.\ Its, B.-Q.\ Jin, and V.E.\ Korepin, 
        % Entanglement in the XY spin chain. 
        J.\ Phys.\ A {\bf 38}, 2975 (2005);
        M.B.\ Plenio, J.\ Eisert, J.\ Dreissig, and M.\ Cramer,
        Phys.\ Rev.\ Lett.\ {\bf 94}, 060503 (2005);        
        M.M.\ Wolf, F.\ Verstraete, M.B.\ Hastings, and 
        J.I.\ Cirac, arXiv:0704.3906;
        %    Title: Area laws in quantum systems: mutual information and        correlations
        M.B.\ Hastings, arXiv:0705.2024.
        
\bibitem{Schuch}
        F.\ Verstraete, J.I.\ Cirac, 
        Phys.\ Rev.\ B {\bf 73}, 094423 (2006).
                        
\bibitem{Calabrese}
        P.\ Calabrese and J.\ Cardy,
        J.\ Stat.\ Mech.\ 0504, P010 (2005);
        J.\ Eisert and T.J.\ Osborne,
        Phys.\ Rev.\ Lett.\ {\bf 97}, 150404 (2006);
        S.\ Bravyi, M.B.\ Hastings, and F.\ Verstraete,
        ibid.\ {\bf 97}, 050401 (2006).

\bibitem{MERA}
        G.\ Vidal, Phys.\ Rev.\ Lett.\ {\bf 99}, 220405 (2007)
        arXiv:quant-ph/0610099;
        M.\ Stewart-Siu and M.\ Weinstein, 
        arXiv:cond-mat/0608042;        
        M.\ Rizzi, S.\ Montangero, 
        arXiv:0706.0868.
        
\bibitem{Flow}
        C.M.\ Dawson, J.\ Eisert, and T.J.\ Osborne,
        arXiv:0705.3456.

\bibitem{Ve04b}
        F.\ Verstraete and J.I.\ Cirac,  
        arXiv:cond-mat/0407066; 
        V.\ Murg, F.\ Verstraete, and J.I.\ Cirac, 
        Phys.\ Rev.\  A \textbf{75},
        033605 (2007).
        
\bibitem{Du05} 
        W. D\"ur et al., Phys.\ Rev.\ Lett.\ \textbf{94},
        097203 (2005).

\bibitem{Ca05}
        J.\ Calsamiglia 
        et al., % L. Hartmann, W. D\"ur and H.J. Briegel, 
        Phys.\ Rev.\ Lett.\ {\bf 95}, 180502 (2005).
%``Spin gases: Quantum entanglement driven by classical kinematics''

\bibitem{Weighted} 
        S.\  Anders, M.B.\ Plenio, W.\ D{\"u}r, 
        F.\ Verstraete, and H.J.\ Briegel, 
        Phys.\ Rev.\ 
        Lett.\  {\bf 97}, 107206 (2006).
          
\bibitem{Weighted1}
        S.\ Anders, H.J.\ 
        Briegel, and W.\ D{\"u}r, New J.\ Phys.\ 
        {\bf 9}, 361 (2007).

\bibitem{Biham} 
        Y.\ Most, Y.\ Shimoni, and O.\ Biham,
        Phys.\ Rev.\ A {\bf 76}, 022328 (2007).
             
\bibitem{Kr07}
        C.\ Mora and H.J.\ Briegel, 
        Phys.\ Rev.\ Lett.\ {\bf 95}, 200503 (2005);
        F.\ Benatti et al., 
        Commun.\ Math.\ Phys.\ {\bf 265},  437 (2006).       
\end{thebibliography}
\end{document}